\documentclass{article}
\usepackage[utf8]{inputenc}
\def\uppaal{{\scshape Uppaal}}

\def\stateflow{Stateflow}
\def\flowchart{flowchart}

\def\sesf{{\scshape SESf}}

\usepackage{cite}
\usepackage{graphicx}
\usepackage{hyperref}
\usepackage{mathtools}
\usepackage{listings}
\usepackage{multirow,tabularx}
\usepackage{verbatim}
\usepackage{caption}
\usepackage{subcaption}
\usepackage{float}
\usepackage{amssymb,amsmath,amsthm,enumitem}
\usepackage[toc,page]{appendix}
\usepackage{styles/z3/style}
\usepackage{styles/proof}
\usepackage{xcolor}
\usepackage{breqn}
\usepackage{textcomp}
\usepackage{pgfplots}
\pgfplotsset{compat=1.14}

\title{Bounded Invariant Checking \\ for Stateflow Programs}
\author{
    Predrag Filipovikj\thanks{Corresponding author.} \quad Dilian Gurov \\ 
    {\small KTH Royal Institute of Technology}\\
    {\small Stockholm, Sweden}\\
    {\small \texttt{\{fpredrag, dilian\}@kth.se}}
    \and 
    Mattias Nyberg \\
    {\small Scania AB CV}\\
    {\small Stockholm, Sweden}\\
    {\small \texttt{mattias.nyberg@scania.com}}
}

\newcolumntype{s}{>{\hsize=.25\textwidth}X}

\newcommand{\bexpr}[1]{\mathcal{SB}[\![#1]\!]}
\newcommand{\aexpr}[1]{\mathcal{SA}[\![#1]\!]}

\newtheorem{theorem}{Theorem}
\newtheorem{proposition}{Proposition}
\newtheorem{definition}{Definition}

\begin{document}
\date{}
\sloppy

\maketitle

\begin{abstract}
\stateflow{} models are complex software models, often used as part of safety-critical software solutions designed with Matlab Simulink. They incorporate design principles that are typically very hard to verify formally. In particular, the standard exhaustive formal verification techniques are unlikely to scale well for the complex designs that are developed in industry. Furthermore, the \stateflow{} language lacks a formal semantics, which additionally hinders the formal analysis.

To address these challenges, we lay here the foundations of a scalable technique for provably correct formal analysis of \stateflow{} models, with respect to invariant properties, based on bounded model checking (BMC) over symbolic executions. The crux of our technique is: i)~a representation of the state space of \stateflow{} models as a symbolic transition system (STS) over the symbolic configurations of the model, as the basis for BMC, and ii)~application of incremental BMC, to generate verification results after each unrolling of the next-state relation of the transition system. To this end, we develop a symbolic structural operational semantics (SSOS) for \stateflow{}, starting from an existing structural operational semantics (SOS), and show the preservation of invariant properties between the two. Next, we define bounded invariant checking for STS over symbolic configurations as a satisfiability problem. We develop an automated procedure for generating the initial and next-state predicates of the STS, and propose an encoding scheme of the bounded invariant checking problem as a set of constraints, ready for automated analysis with standard, off-the-shelf satisfiability solvers. Finally, we present preliminary performance results by applying our tool on an illustrative example.
\end{abstract}
\section{Introduction}
\label{sec:introduction}

\stateflow{}~\cite{stateflowug} is a proprietary graphical modelling language developed and maintained by Mathworks. It is an extension of a formalism for modelling complex systems through hierarchical state machines called Statecharts~\cite{harel1987statecharts}. The expansion and popularity of the \stateflow{} language is mostly due to its integration into the Matlab Simulink graphical development and simulation environment~\cite{simulinkug}, which makes it a standard tool for modeling (hierarchical) state-machines in many industrial domains, such as the automotive, avionics, and railway domains. The rich graphical formalism and the variety of supporting tools in the Matlab Simulink environment enable the development of highly complex software models, which in many instances are classified as \emph{safety-critical}. The correctness of safety critical systems is regulated by domain-specific safety standards (e.g., ISO26262~\cite{iso26262} in the automotive domain), which require correct operation of such systems at all times with strongly regulated error margins. 

One way of enabling a high level of quality-assurance for safety-critical systems is to employ rigorous mathematics-based verification methods, popularly known as \emph{formal verification techniques}. The starting point of applying such techniques is the existence of a precise \emph{formal model} of the system design. To be able to generate such a model, besides a formal \emph{syntax}, the modeling language in which the system design is produced must have a precisely defined execution \emph{semantics}. However, even in its most comprehensive documentation~\cite{stateflowug}, the execution semantics of the \stateflow{} language has only been described informally. Still, Mathworks offers a tool, namely the Simulink Design Verifier~(SLDV)~\cite{stateflowug}, which provides features for test case generation and formal property checking. The SLDV tool is of a closed-source, proprietary nature, and not much information about the employed formal verification technique has been disclosed so far, except that it implements SAT-based model checking~\cite{hamon2008simulink}. According to Etienne et al.~\cite{etienne10usingsldv}, SLDV implements SAT-based bounded model checking and $k$-induction using the Prover SL DE tool~\cite{abdulla2004scadedesignverifier}, yet little information is presented as to how exactly the original \stateflow{} models are analysed. On top of the information scarcity, the SLDV tool is distributed under a license that explicitly forbids benchmarking or any other form of direct comparison with another approach or tool, be it commercial or of purely academic nature.

The problem of formal verification of \stateflow{} models has been addressed in a number of research endeavours, which have focused either on defining a formal semantics for the language~\cite{hamon2004operationalsemanticsstateflow,hamon2007operationalsemanticsstateflow,hamon2005denotational,bourbouh2017automatedanalysisofstateflowmodels}, aimed to be used as the basis for formal analysis, or propose a model-to-model transformation scheme of \stateflow{} models into some formal analysis framework, without having an underpinning formal semantics for \stateflow{} itself~\cite{yang16stateflowtransformationase,jiang2019dependablecpsstateflow,banphawatthanarak1999symbolic}. The \stateflow{} verification approaches predominantly focus on \emph{exhaustive} techniques, in order to determine whether a \stateflow{} model satisfies given property or not. However, due to the complex nature of the \stateflow{} models, particularly in industrial settings, generating a formal correctness certificate is often intractable, mostly due to the state-space explosion problem~\cite[p.~77]{katoen08principlesofmc}. In such cases, the verification task runs out of physical memory and fails, without providing much useful feedback to the designers. 
Apart from this \emph{scalability} limitation, some of the approaches for model-to-model transformation (see e.g.~\cite{jiang2019dependablecpsstateflow}) lack the means to formally demonstrate the correctness of the formal analysis model w.r.t. the original \stateflow{} model, and consequently of the obtained analysis results. 

In this work, we are tackling the aforementioned challenges for formal analysis of \stateflow{} models by presenting a technique that applies \emph{bounded model checking} (BMC)~\cite{biere2003boundedmodelchecking} over \emph{symbolic executions}~\cite{king1976symbolicexecutionandtesting} of \stateflow{} models. We adopt BMC as the underlying technique for verification for two main reasons: first, to leverage the power of SAT/SMT-based model checking~\cite{barrett18smtbookchapter}, and second, to alleviate the state-space explosion by incrementally exploring all system executions of bounded length~\cite{barrett18smtbookchapter}, until the problem becomes intractable. In this paper, we focus on checking \emph{invariant} properties, which are state properties that hold in all reachable states of a given program. Even though invariant properties represent just one class of properties, based on our previous and current experiences in collaboration with industrial partners, it is often considered to be the most important one for safety-critical systems.

There are two crucial aspects of our verification technique. Fist, we develop a provably correct transformation of \stateflow{} models into \emph{symbolic transition systems} (STS), and second, we encode the invariant checking problem for STS as a quantifier-free SMT problem that can be checked using any modern SMT solver. 

\paragraph{Contributions}
Our verification technique consists of the following ingredients. First, we derive a set of symbolic structural operational semantics rules (SSOS). The SSOS rules are obtained by uniformly translating into symbolic counterparts the rules of an already existing SOS for \stateflow{}~\cite{hamon2004operationalsemanticsstateflow}. We build on top of this particular set of SOS rules, because it is the only available operational semantics for \stateflow{} that is suitable for our needs, and because the correctness of the rules has already been validated against the \emph{simulation semantics} of \stateflow{} (see~\cite{hamon2004operationalsemanticsstateflow}). The SSOS is needed for deriving STS at a suitably high level of granularity of the execution steps (which we choose to be the level of \stateflow{} program statements), abstracting from the intricate many-layered transitions of the original SOS. 
As our second contribution, we present two theorems that show that the SOS and SSOS \emph{simulate} each other. This result is crucial for the correctness of our technique. 
Our third contribution is a translation, using the SSOS, of \stateflow{} programs into STS over symbolic configurations, and the encoding of the STS and the given invariant property into a set of constraints in the SMT-LIB format~\cite{barret15smtlib}. The latter set of constraints can then be used as input to most of the modern SMT solvers. In our work, we use the Z3 SMT solver~\cite{demoura08z3} from Microsoft Research. Finally, we present preliminary evidence for the practical usefulness of our approach, by applying it on an illustrative \stateflow{} model. Even though initially we planned to compare our approach against the SLDV tool, in the end it was not possible due to the strict licensing constraints imposed by Mathworks.

\paragraph{Structure}
Our paper is organised as follows. In Section~\ref{sec:background} we outline the required background concepts that we use throughout the paper, including a brief overview of the \stateflow{} modelling language illustrated on a running example (Section~\ref{sec:stateflow}), the imperative \stateflow{} language and its formal execution semantics (Section~\ref{sec:h-r-sos}), and an overview of the SMT (Section~\ref{sec:smt-z3}) and BMC (Section~\ref{sec:bmc}) techniques. In Section~\ref{sec:symbolicsos} we present the SSOS for \stateflow{} programs, followed by the characterization of the relationship between the concrete and symbolic semantics in Section~\ref{sec:characterization-sos-ssos}. Next, in Section~\ref{sec:transformation-and-smt-encoding}, we show how an STS over symbolic configurations can be constructed using the SSOS rules (Section~\ref{sec:stateflow-to-sts}), followed by an informal encoding procedure into an SMT-LIB script (Section~\ref{sec:sts-to-smt}). Next, we show a preliminary evaluation of our approach, based on the running example (Section~\ref{sec:evaluation-and-comparison}). We give an overview of the related work in Section~\ref{sec:related-work}, and finally, in Section~\ref{sec:conclusions}, present our conclusions and outline some directions for future work. 

\section{Background}
\label{sec:background}

In this section, we present an overview of the concepts on which we build our work. First, in Section~\ref{sec:stateflow} we give a succinct overview of the \stateflow{} modeling language. Next, in Section~\ref{sec:h-r-sos} we give a brief overview of the existing \stateflow{} imperative language and its SOS. In Section~\ref{sec:smt-z3} we recall the general concept of Satisfiability Modulo Theories (SMT) and the Z3 tool, and finally in Section~\ref{sec:bmc} we give an overview of Bounded Model Checking (BMC).

\subsection{\stateflow{}}
\label{sec:stateflow}

\begin{figure}[t!]
\centering 
    \includegraphics[width=\textwidth]{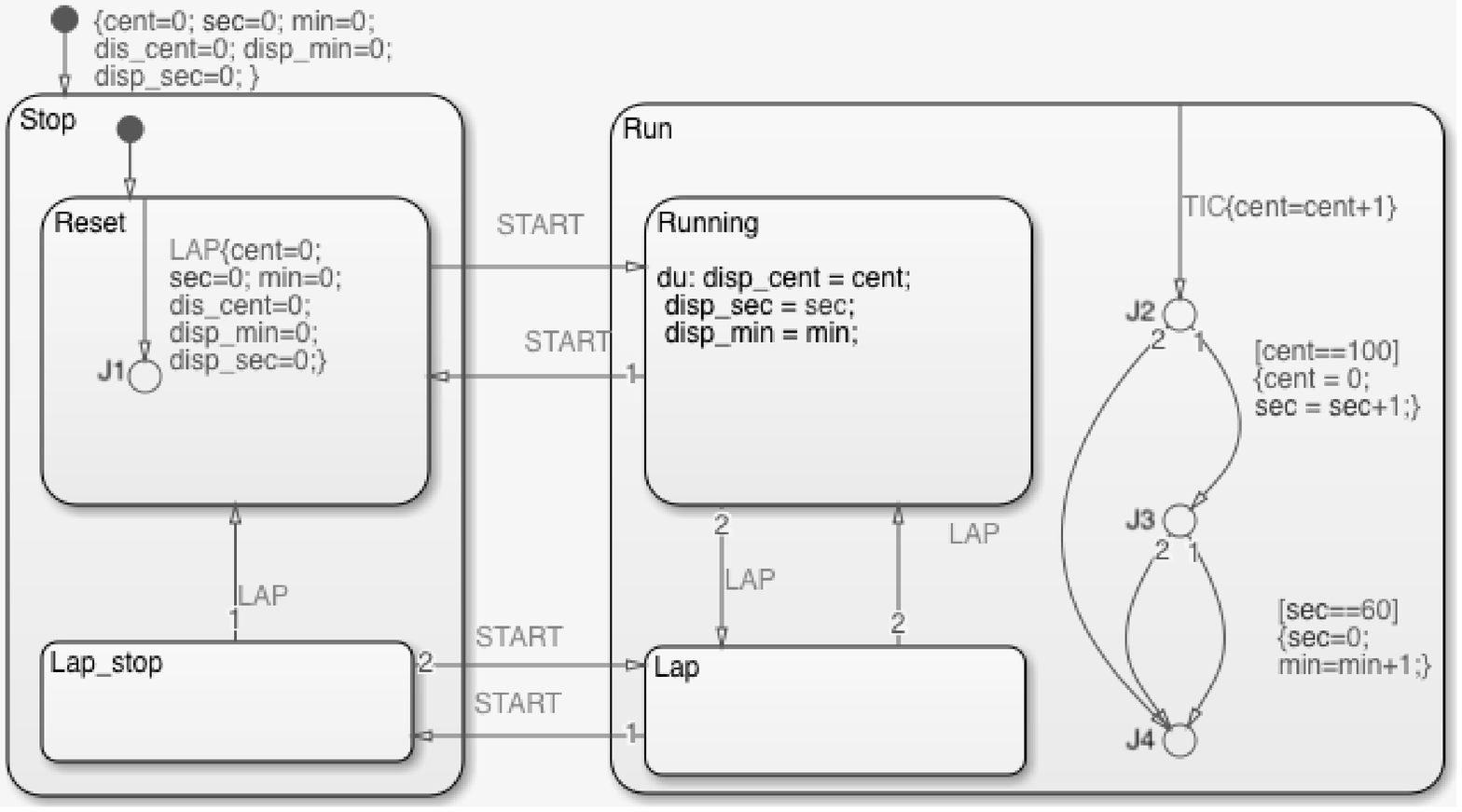} \caption{Simple \stateflow{} diagram - timer example~\cite{hamon2004operationalsemanticsstateflow}.}
    \label{fig:stateflow-scenario}
\end{figure}

\stateflow{}~\cite{stateflowug} is a graphical modeling language developed by Mathworks, integrated into the Matlab Simulink~\cite{simulinkug} modelling environment. The language provides means for modelling of hierarchical state machines through the following features: state transition diagrams, \flowchart{}s, state transition tables, and truth tables. In our work, we focus on the state transition diagrams and \flowchart{}s.

A Simulink \stateflow{} diagram can be broadly divided into two parts: \emph{control} and \emph{data}. The control part is modeled through the concepts of \emph{Stateflow state}, \emph{connective junction}, and \emph{transition}, whereas the data part is modelled through a set of \emph{data variables} and \emph{events}. 

The control of the \stateflow{} diagram in Figure~\ref{fig:stateflow-scenario} consists of 6 Stateflow states, 4 connective junctions and 13 transitions. We have chosen the particular model as a running example because it has relatively simple control flow that helps us to illustrate our concepts, yet it yields an infinite state-space due to the unbounded integer variable \emph{min}. Each \stateflow{} state is decorated with a set of \emph{state actions}. The basic set of state actions (denoted as $A$) includes: \emph{entry ($en$)}, \emph{duration ($du$)} and \emph{exit ($ex$)}. Each action represents an atomic routine. A Stateflow state is either \emph{atomic} or \emph{composite}. Composite Stateflow states contain other states (called \emph{substates}) in their internal structure. Based on how their substates are composed, the composite states can be either \texttt{Or} or \texttt{And}. In \texttt{Or}-compositions there can be only one active substate, whereas in \texttt{And}-compositions, all of the substates are active in the same time. The parallelism in the context of \texttt{And}-compositions only means concurrent activation of its substates; the execution, however, is strictly sequential, according to a predefined order. \emph{Connective junction} (or simply \emph{junction} for short), is used for modelling different branches of execution when a \stateflow{} diagram moves from one control point into another. The junctions are represented as hollow circles (\emph{J1} to \emph{J4} in Figure~\ref{fig:stateflow-scenario}).

The dynamics of the control flow of a \stateflow{} diagram is modelled through a set of \textit{transitions} of the following format: $\mathit{s \xrightarrow{e, c, ca, ta} s'}$, where $s$ and $s'$ are the \emph{source} and the \emph{destination} state or junction, respectively, $e$ is the \emph{transition event} upon which a transition fires (we use the term \emph{``fires"} as a synonym for executing the transition), the firing of a transition is enabled by its \emph{transition condition} $c$; $\mathit{ca}$ is the \emph{conditional action} that is executed if the condition evaluates to true and before the transition execution is completed, and finally $\mathit{ta}$ is a \emph{transition action} that is performed after the transition execution is completed. Transitions can be either \emph{full transitions} (both source and destination are states) or \emph{transition segments} (at least source or destination are junction). The control-flow of a \stateflow{} program evolves only through completion of full transitions.
Transitions can be grouped in sets called \emph{transition lists}, based on a common source element. The internal execution of a transition list is strictly sequential following the order assigned at design-time. There are two types of transition lists: \emph{internal} and \emph{outgoing}, with the former being characteristic for states only, whereas the later one for both states and junctions. The action language is separate from the \stateflow{} itself, and the actions can be specified in the following languages: C, C++, Fortran, or Matlab.

The informal execution semantics of \stateflow{} models is very intricate and has been explained in detail in the \stateflow{} user guide published by Mathworks~\cite{stateflowug}. We omit here the details of the informal execution semantics, but give in the following section an overview of a formal one.

\subsection{\stateflow{} Imperative Language: Formal Syntax and Structural Operational Semantics}
\label{sec:h-r-sos}

\begin{table}[!htbp]
\centering
\caption{The Stateflow Imperative Language by Hamon and Rushby.}
\begin{tabular}{*9c}
\multicolumn{4}{l}{composition} 
& \multicolumn{1}{l}{}
& \multicolumn{4}{l}{$C=Or(s_a,p,T,SD) \, \vert \, And(b, SD)$}\\
\multicolumn{4}{l}{state definition} 
& \multicolumn{1}{l}{}
& \multicolumn{4}{l}{$sd = ((a,a,a), C, T_i, T_o, J)$}\\
\multicolumn{4}{l}{state definition list} 
& \multicolumn{1}{l}{}
& \multicolumn{4}{l}{$SD = \{s_0:sd_0, \dots, s_n:sd_n \}$}\\
\multicolumn{4}{l}{junction definition list} 
& \multicolumn{1}{l}{}
& \multicolumn{4}{l}{$J = \{j_0:T_0, \dots, j_n:T_n \}$}\\
\multicolumn{9}{l}{}\\
\multicolumn{2}{l}{transition list} 
& \multicolumn{3}{l}{}
& \multicolumn{4}{l}{$T= \emptyset_T \, \vert \, t.T$}\\
\multicolumn{2}{l}{transition} 
& \multicolumn{3}{l}{}
& \multicolumn{4}{l}{$t = (e_t, c, a, a, d)$}\\
\multicolumn{1}{l}{state} 
& \multicolumn{4}{l}{$s$}
& \multicolumn{1}{l}{active state}
& \multicolumn{1}{l}{}
& \multicolumn{2}{l}{$s_a = \emptyset_s \, \vert \, s$}\\
\multicolumn{1}{l}{junction} 
& \multicolumn{4}{l}{$j$}
& \multicolumn{1}{l}{}
& \multicolumn{1}{l}{}
& \multicolumn{2}{l}{}\\
\multicolumn{1}{l}{path} 
& \multicolumn{4}{l}{$p = \emptyset_p \, \vert \, p$}
& \multicolumn{1}{l}{destination}
& \multicolumn{1}{l}{}
& \multicolumn{2}{l}{$d = p \, \vert \, j$}\\
\multicolumn{1}{l}{event} 
& \multicolumn{4}{l}{$e$}
& \multicolumn{1}{l}{transition event}
& \multicolumn{1}{l}{}
& \multicolumn{2}{l}{$e_t = \emptyset_e \, \vert \, e$}\\
\multicolumn{1}{l}{action} 
& \multicolumn{4}{l}{$a$}
& \multicolumn{1}{l}{condition}
& \multicolumn{1}{l}{}
& \multicolumn{2}{l}{$c$}\\
\end{tabular}
\label{tab:imperative-language}
\end{table}

In order to formalize \stateflow{}, Hamon and Rushby propose in~\cite{hamon2004operationalsemanticsstateflow,hamon2007operationalsemanticsstateflow} an imperative language that is a strict subset of the \stateflow{} graphical language. To the best of our knowledge, the imperative language supports most of the graphical language features except for the history nodes, and as such is expressive enough for most of the industrial \stateflow{} models. A \emph{\stateflow{} program} is simply a \stateflow{} model (re)written in the imperative language. The execution of \stateflow{} programs is formalised by a set of 27 rules presented in a structural operational semantics (SOS) style. 
In the following, we give a brief overview of the language and its operational semantics.

The syntax of the \stateflow{} imperative language is given in Table~\ref{tab:imperative-language}. The language is based on the following syntactic categories: state ($s$), junction ($j$), event ($e$), action ($a$) and condition ($c$). A transition $t = (e_t, c, a_c, a_t, d)$ is composed of a transition event $e_t$, condition $c$, condition and transition actions $a_c$, $a_t$, respectively, and a destination~$d$ to which it fires. Transitions are grouped into transition lists. A junction definition list~$J$ associates a list of transitions with junctions. A state definition list $SD$ associates each state variable ($s$) with a state definition $sd = ((a,a,a), C, T_i, T_o, J$). Each $sd$ contains 3 actions, a composition~$C$, lists of internal and outgoing transitions $T_i$ and $T_o$, respectively, and a junction definition list $J$. Finally, the composition~$C$ can be of type $\mathtt{Or}(s_a, p, T, SD)$, where $s_a$ is the active state, $p$ is the path, $T$ is a transition list, and $SD$ is a list of state definitions; or of type $\mathtt{And}(b, SD)$, which has a Boolean value~$b$ signifying whether the component is active or not, a path~$p$, and a state definition list~$SD$. In the reminder of the manuscript, we will use the term \textit{\stateflow{} program} regardless if it is modeled using the original \stateflow{} graphical language or the imperative language as we are going to be handling only models that can be rewritten in the imperative language.

The execution of a \stateflow{} program consists of processing an input event through a sequence of discrete steps. The execution involves modifying the value of the program variables, raising output events, and changing the program control point by executing transitions. The operational semantics is formalised by a set of 27 layered rules, which precisely prescribe the sequence of actions involved in the processing of an event through the elements of the imperative language. The operational semantics has been validated against the simulation semantics of the Matlab \stateflow{} simulation engine by performing extensive simulations with \stateflow{} graphical models and comparing the output traces with the (SOS based) executions of the corresponding \stateflow{} programs~\cite{hamon2004operationalsemanticsstateflow}. In this work, we refer to executions derivable using the SOS rules as \emph{concrete}. The following judgment expresses the base form of event processing in \stateflow{} programs~\cite{hamon2004operationalsemanticsstateflow}:

\begin{equation*}
    \mathit{e \vdash (P, D) \rightarrow (P', D'), tv}
\end{equation*}

\noindent which reads as follows: processing an event $e$ in an environment $D$ through a program component $P$ produces a new environment $D'$, a new program component $P'$ and a transition value $\mathit{tv}$. An environment $\mathit{D: Var \rightarrow Val}$ is a mapping from variables to values. $Env$ denotes the set of all possible environments; $P$ is an element of the \stateflow{} imperative language, whereas $\mathit{tv \in \{Fire(d,a) \: \vert \: No \: \vert \: End\}}$ is a transition value which indicates whether a transition has fired ($\mathit{Fire(d,a)}$) or not ($\mathit{No \: \vert \: End}$). All of the rules in the SOS extend and slightly differ from this general form~\cite{hamon2007operationalsemanticsstateflow}.

\begin{figure*}[t!]
    \centering
\begin{subfigure}[b]{.99\textwidth}
    \centering
    \begin{minipage}{.99\textwidth}
    \infer{e \vdash ((e_0, c, ca, ta, d), D_1) \rightarrow{} D_2, Fire(d, ta)}{(e = e_0) \lor (e_0 = \emptyset{}) & e \vdash (c, D_1) \rightarrow{} \top & e \vdash (ca, D_1) \hookrightarrow D_2}
    \end{minipage}
    \caption{[t-FIRE]\textsubscript{SOS}  rule~\cite{hamon2007operationalsemanticsstateflow}}
    \label{fig:t-fire-sos-rule}
    \end{subfigure}%
\\~\\
\begin{subfigure}[b]{.99\textwidth}
    \centering
    \begin{minipage}{.5\textwidth}
        \infer{e \vdash ((e_0, c, ca, ta, d), \langle\Delta_1, pc_1 \rangle) \rightarrow{} \langle\Delta_2, pc_2 \rangle, Fire(d,  ta)}{
        \begin{array}{lc}
            (e = e_0) \lor (e_0 = \emptyset{}) \\
            e \vdash (c, \langle\Delta_1, pc_1 \rangle) \rightarrow{} \langle\Delta_1, pc_2 \rangle & e \vdash (ca, \langle\Delta_1, pc_2 \rangle) \hookrightarrow \langle\Delta_2, pc_2 \rangle 
        \end{array}}
    \end{minipage}
\caption{[t-FIRE]\textsubscript{SSOS} rule}
\label{fig:t-fire-ssos-rule}
\end{subfigure}%
\caption{The [t-FIRE] rule in (a) SOS and (b) SSOS semantics.}
\end{figure*}

The [t-FIRE] rule for both SOS and SSOS semantics is shown in Figure~\ref{fig:t-fire-sos-rule} and Figure~\ref{fig:t-fire-ssos-rule}, respectively. The rule describes how a \stateflow{} transition fires, and intuitively it captures the following: in the concrete execution, if the evaluation of a condition evaluates to true ($\top$), and the execution of the condition action $ca$ modifies the environment, then a \stateflow{} program performs a transition, and raises a \emph{Fire} transition value. In Figure~\ref{fig:t-fire-ssos-rule}, we show an SSOS counter-part of the [t-FIRE] rule to visually illustrate the similarities and differences of the rules side-by-side. An intuitive explanation for the rule is given in Section
~\ref{sec:symbolicsos}. 
For the complete set of SOS rules, we refer the interested reader to the original work by Hamon~and~Rushby~\cite{hamon2004operationalsemanticsstateflow,hamon2007operationalsemanticsstateflow}.

\subsection{Satisfiability Modulo Theories and Z3}
\label{sec:smt-z3}

The problem of determining whether a Boolean formula can be made true by assigning truth values to the constituent Boolean variables is known as the \emph{Boolean satisfiability problem} (SAT). A decision procedure for SAT is a procedure that generates a (satisfying) assignment for the variables for which a given formula is true, whenever the formula is satisfiable. \emph{Satisfiability Modulo Theories} (SMT) represents an extension of SAT, where some of the logic symbols are interpreted by a background theory~\cite{barrett18smtbookchapter}. An example of such a background modulo-theories are the theory equality, theory of integer numbers, theory of real numbers, etc. 

Z3~\cite{demoura08z3} is a state of the art SMT solver and theorem prover developed by Microsoft Research. The input is a model specified in a text-based assertion language that follows the SMT-LIB standard~\cite{barret15smtlib}. Z3 provides a number of APIs for different programming languages, including C and Python, which enables the integration of the Z3 solver with other applications. The input model consists of a set of variables of specific types (also called sorts), and a set of assertions that express constraints over the variables. The basic command called \texttt{assert} is used to add an assertion to the internal stack of the solver. Once the stack is loaded with the set of assertions of interest, the satisfiability of the constraints is checked by the \texttt{check-sat} command. There are 3 possible outcomes from the decision procedure: \texttt{sat} which indicates that there exists an assignment of the variables satisfying the set of assertions, \texttt{unsat} indicating that there does not exist such a satisfying assignment, and \texttt{unknown} when the decision procedure cannot determine the satisfaction of the assertions on the stack.

\subsection{Bounded Model Checking}
\label{sec:bmc}

In this section, we give an overview of the Bounded Model Checking (BMC) technique, which we use for checking invariant properties of \stateflow{} programs.
    
BMC is a refutation-based verification technique, in which a symbolic representation of the system behavior is unrolled for a predefined number~$k$ of steps, called the \emph{reachability diameter}. It has been shown that checking a property over a finite set of states can be reduced to checking the satisfiability of a corresponding propositional formula. Thus, the goal of BMC is to generate a formula that is satisfiable if there exists a violation of the property in some execution of length up to~$k$. The reduction of the model-checking problem to a satisfiability problem is motivated by the increase in the computational power of modern solvers, which tend to be more efficient in solving large formulas as compared to techniques based on BDDs~\cite{demoura03BMCrefutationtoverification}.

To be able to formally define the BMC problem for invariant properties, we first recall some additional background concepts. A common way of capturing program behavior is via a \emph{transition system} (TS), formally defined as follows.
\begin{definition}[Transition System]
\label{def:path-ts}
A \emph{transition system} is a tuple $\mathit{TS} = (C, C_0, \rightarrow)$, where: $C$ is a finite set of \emph{configurations}, $C_0 \subseteq C$ a set of \emph{initial configurations}, and $\mathit{\rightarrow \subseteq C \times C}$ a \emph{transition relation}.  

Let $\pi$ be a finite or infinite sequence of configurations, written $\pi = c_0, c_1, \ldots, c_n$ or $\pi = c_0, c_1, c_2 \ldots$, respectively. The sequence $\pi$ is called:
\begin{itemize}
    \item a \emph{path} of $\mathit{TS}$, written path($\pi$), if $\forall c_i, c_{i+1} \in \pi.\ (c_i, c_{i+1}) \in \rightarrow$, and
    \item an \emph{initialized path}, if path($\pi$) and $c_0 \in C_0$.
\end{itemize}
A finite path $\pi = c_0, c_1, \ldots, c_n$ is said to be of \emph{length}~$n$. 
\end{definition}

Transition systems are a useful concept for reasoning about the \emph{concrete computations} of programs. A configuration $\mathit{c \in C}$ of a program is a pair $\mathit{c = (l, v)}$, where $l$~is a control point in the program, and $v$~is a mapping between the program variables and values taken from their respective domains. 

When programs contain variables that range over infinite domains, it can be more efficient to reason about their behaviors in a \emph{symbolic} way, through a set of predicates. This gives rise to the notion of symbolic transition system (STS).

\begin{definition}[Symbolic Transition System]
\label{def:sts-with-predicates}
A \emph{symbolic transition system} is a pair $S = (I, R)$, where the unary predicate~$I(\cdot)$ is a first-order logic (FOL) formula over the components of configurations representing the initial set of configurations, and the binary predicate~$R(\cdot, \cdot)$ is a formula representing the ``next-state" transition relation, satisfying the equivalences:
\begin{align*}
    &\mathit{I(c) \: \Leftrightarrow c \in C_0}\\
    &\mathit{R(c,c')\: \Leftrightarrow (c,c') \in \ \rightarrow}
\end{align*}
\end{definition}

Every initialized path in $\mathit{S}$ of length~$k$ can be characterized by the formula:
\begin{equation}
\label{eq:path-encoding}
    \mathit{path} (c_0, c_1, \ldots, c_k) \: \triangleq \: I(c_0) \land \bigwedge\limits_{i=0}^{k - 1} R(c_i,c_{i+1}), 
\end{equation}
and then, the existence of an initialized path of length~$k$ is equivalent to the satisfiability of the formula $\mathit{path} (x_0, x_1, \ldots, x_k)$, where $x_i$ is a variable representing a configuration.

Since BMC operates over a subset of the reachable configurations, contained within the given reachability diameter~$k$, in this work we use the term \emph{$k$-bounded invariant property} to denote an invariant property that holds over the reachability diameter~$k$. 

Let $\varphi$ be a unary predicate over configurations, i.e., a  property. We define the corresponding $k$-bounded invariant property, denoted~$\varphi^k$, as the formula:

\begin{equation}
\label{eq:k-bounded-invariance-definition}
   \forall c_0, c_1, \dots, c_k.\ (\mathit{path} (c_0, c_1, \dots, c_k) \Rightarrow \bigwedge\limits_{i=0}^{k} \varphi (c_i))
\end{equation}~

\noindent
To disprove such a $k$-bounded invariant property, it is sufficient to show that there exists a configuration within the reachability diameter for which $\varphi^k$~does not hold. A path containing such a configuration is called a \emph{counter-example}, and is characterized by the logical negation of the above formula, i.e.:

\begin{equation}
\label{eq:counter-example-definition}
    \exists c_0, c_1, \dots, c_k.\ (path(c_0, c_1, \dots c_k) \land 
    \bigvee\limits_{i=0}^{k} \neg \varphi (c_i))
\end{equation}~

Given that the predicates $I$, $R$, and $\varphi$ can be expressed as FOL formulas, where some function and predicate symbols are potentially interpreted by some background theory, it is obvious how the refutation of $k$-bounded invariant properties can be reduced to an SMT problem. In case that there exists a satisfying assignment for (\ref{eq:counter-example-definition}), a counter-example for the invariant property is generated. Conversely, if there exists no such satisfying assignment, the $k$-bounded invariant property holds.

\section{Symbolic Structural Operational Semantics}
\label{sec:symbolicsos}

In this section, we present our SSOS semantics for the \stateflow{} imperative language, which we use as a basis for constructing an STS $\mathit{\widehat{S}}$ for a given \stateflow{} program. We start from the existing SOS semantics as in~\cite{hamon2004operationalsemanticsstateflow,hamon2007operationalsemanticsstateflow}, and transform each of the SOS rules uniformly in a corresponding symbolic counterpart.

From earlier (see Section \ref{sec:stateflow}), we know that the SOS rules for \stateflow{} programs are over judgments of the following form:

\begin{equation*}
   \mathit{e \vdash (P, D) \rightarrow (P', D'), tv}
\end{equation*}~

Based on the set of SOS rules, one can induce a $TS = (C, C_0, \rightarrow)$, where $C$ is the set of concrete configurations, each configuration $c \in C$ being a tuple $c = (P, D)$, $C_0 \subseteq C$ is the set of initial configurations, and $\rightarrow \subset C \times C$ is the transition relation, composed of transitions derivable by the SOS rules.

In the original formalization, the sets of variables ($\mathit{Var}$) and values ($\mathit{Val}$), as well as the sets of actions ($\mathit{Act}$) and conditions ($\mathit{Cond}$) are considered to be a part of the action language which is distinct from the \stateflow{} language itself. The details for the actions and conditions are abstracted away; however, it is assumed that the semantics of the executing actions and the evaluating conditions is available via judgments of the form: 

\begin{equation*}
    \mathit{\textnormal{(i) } e \vdash (a, D) \hookrightarrow{} D' \textnormal{ and }\textnormal{(ii) } e \vdash (c, D) \rightarrow \top \: \vert \: \bot} 
\end{equation*} 
~

\noindent
which are read as follows: (i) evaluating an action ($a$) in a current environment ($D$) produces a new environment ($D'$), and (ii) evaluating a condition ($c$) in an environment ($D$) produces either true or false Boolean value.

The set of SSOS rules is created by uniformly transforming each of the SOS rules into a corresponding symbolic rule, by: i) replacing each valuation of the program variables, called \emph{environment} ($\mathit{D}$) with a symbolic representation ($\Delta$), and ii) adding a path condition ($\mathit{pc}$). Consequently, we update the action execution and condition evaluation, which evaluate over the symbolic environment and path condition, respectively. Following the basic principles of symbolic execution~\cite{king1976symbolicexecutionandtesting}, in the set of SSOS rules we treat the data component of the language in a symbolic way, whereas the control-flow remains concrete.

As we already discussed in Section~\ref{sec:bmc}, the set of all initialized paths of length \emph{k} of a \stateflow{} program can be symbolically encoded as formulas. In such a representation, the analysis operates over sets of environments, rather than with each environment individually. Conceptually, $\Delta$~represents a set of concrete environments that is characterized by a path formula that includes all concrete environments reachable at a particular point of execution, while $\mathit{pc}$ represents the path condition that encodes the conditions over data such that $\Delta$ is valid.

\begin{figure*}[t!]
\begin{subfigure}[c]{.99\textwidth}
    \centering
    \begin{minipage}{.90\columnwidth}
    \begin{align*}
        \infer{e, J \vdash (t.T, \langle\Delta_1, pc_1 \rangle) \rightarrow{} \langle\Delta_2, pc_2\rangle, Fire(d, ta)}{e \vdash (t, \langle\Delta_1, pc_1 \rangle) \rightarrow{} \langle\Delta_2, pc_2\rangle, Fire(d, ta)}
        \end{align*}
    \end{minipage}
    \caption{[T-FIRE]\textsubscript{SSOS} rule}
    \end{subfigure}%
\\
\begin{subfigure}[c]{.99\textwidth}
    \centering
    \begin{minipage}{.90\columnwidth}
    {
        \begin{align*}
            \infer{\begin{array}{cr}
                  e, J_0 \vdash ((A,C,T_i,T_o,J), \langle\Delta_1, pc_1 \rangle) \rightarrow{} ((A,C',T_i,T_o,J), \langle\Delta_5, pc_3 \rangle, \\
                  Fire(d, \diamond))
            \end{array}}{\begin{array}{lcl}
                 e, J_0 \vdash (T_o, \langle\Delta_1, pc_1 \rangle) \rightarrow{} \langle\Delta_2, pc_2\rangle, Fire(d, ta)\\
                  e \vdash (ta, \langle\Delta_2, pc_2) \hookrightarrow{}\langle\Delta_3, pc_2\rangle \quad e \vdash (C, \langle\Delta_3, pc_2\rangle) \Downarrow (C', \langle\Delta_4, pc_3\rangle)\\
                  e \vdash (A.ex, \langle\Delta_4, pc_3\rangle) \hookrightarrow{} \langle\Delta_5, pc_3\rangle
            \end{array}}
        \end{align*}
    }
    \end{minipage}
\caption{[SD-FIRE]\textsubscript{SSOS} rule}
\end{subfigure}%
\\
\begin{subfigure}[c]{.99\textwidth}
    \centering
    \begin{minipage}{.90\columnwidth}
    \begin{align*}
    \infer{\begin{array}{cc}
          e, J, tv \vdash (And\{s_0:sd_0 \cdots s_n:sd_n\}, \langle\Delta_1, pc_1\rangle) \\
         \rightarrow{} (And\{s_0:sd'_0 \cdots s_n:sd'_n\}, \langle\Delta_{n+1}, pc_{n+1}\rangle, No)
    \end{array}}{\begin{array}{lc}
         (tv = No) \lor (tv = End)\\
         \forall i \in [1,\dots, n] \quad e \vdash (sd_i, \langle\Delta_i, pc_i\rangle) \rightarrow{} (sd'_i,\langle\Delta_{i+1}, pc_{i+1}\rangle, No)
    \end{array}}
    \end{align*}
    \end{minipage}
    \caption{[AND]\textsubscript{SSOS} rule}
\end{subfigure}%
\\
\begin{subfigure}[c]{.99\textwidth}
    \centering
    \begin{minipage}{.90\columnwidth}
    \begin{align*}
    \infer{\begin{array}{cc}
        e, J \vdash (Or(s,p,T,SD[s:sd]), \langle\Delta_1, pc_1\rangle, tv) \\
    \rightarrow{} (Or(\emptyset_s,p,T,SD[s:sd']), \langle\Delta_2, pc_2\rangle, Fire(p',a))
    \end{array}}{\begin{array}{lc}
          (tv = No) \: \lor (tv = \: End) \\
          e,J,tv \vdash (sd, \langle\Delta_1, pc_1\rangle)\rightarrow{} (sd', \langle\Delta_2, pc_2\rangle, Fire(p', a)) \quad \neg prefix(p',p)\\
    \end{array}}
    \end{align*}
    \end{minipage}
\caption{[OR-FIRE]\textsubscript{SSOS} rule}
\end{subfigure}%
\caption{Illustrative sample of SSOS rules.}
    \label{fig:illustrative-rules}
\end{figure*}

We define a \emph{symbolic configuration} $\mathit{sc} \in \mathit{SC}$ as a structure $\mathit{( P, \langle \Delta, pc\rangle)}$, where $P$~is any component from the imperative language from Table~\ref{tab:imperative-language}. We introduce a new set of \emph{symbolic variables} (symbols), denoted $\mathit{Sym}$, and a bijection $\mathit{g: Var \rightarrow Sym}$ between the program variables and the symbols. The path condition $pc$ is simply a Boolean expression over the set of symbols, whereas the \emph{symbolic environment} $\Delta \in \mathit{SEnv}$ is a mapping $\mathit{\Delta: \mathit{Var} \rightarrow \mathit{Expr_{Sym}}}$ from program variables to (arithmetic) expressions over symbols. Finally, we assume that symbolic action execution and symbolic condition evaluation are provided via semantic functions of type $\mathcal{SA}: \mathit{Act} \rightarrow (\mathit{SEnv} \rightarrow \mathit{SEnv})$ and $\mathcal{SB}: \mathit{Cond} \rightarrow (\mathit{SEnv} \rightarrow \mathit{BExpr_{Sym}})$, respectively.

We can now define the axioms for action execution and condition evaluation, for symbolic execution of \stateflow{} programs, as follows:

\begin{equation}
\hspace*{-3mm}
\begin{aligned}
&e \vdash (a, \langle\Delta_1, pc_1\rangle) \hookrightarrow{} \langle\Delta_2, pc_1\rangle ~~~\textnormal{~if } \Delta_2 = \aexpr{a}(\Delta_1)\\
&e \vdash (c, \langle\Delta_1, pc_1\rangle) \rightarrow \langle\Delta_1, pc_2\rangle ~~~~\textnormal{~if } pc_2 = pc_1 \land \bexpr{c}(\Delta_1)
\end{aligned}
\end{equation}~

\noindent 
The initial symbolic configuration is $(P, \langle pc_0, \Delta_0\rangle)$, where $P$~is a component of the \stateflow{} imperative language, $\Delta_0 = g$, and $pc_0 = \top$.

The set of SOS rules can now be uniformly translated into a corresponding SSOS counter-part. In Figure~\ref{fig:illustrative-rules}, we show a subset of the SSOS rules, in addition to the [t-FIRE] rule from Figure~\ref{fig:t-fire-ssos-rule}. 
The [t-FIRE] rule in Figure~\ref{fig:t-fire-ssos-rule} describes how a \stateflow{} transition ($t$) fires by appending the symbolic evaluation of the condition $t.c$ to the current path condition and by symbolically executing the condition action $t.ca$ over the current symbolic environment $\Delta$. When a transition fires, a transition event $Fire(t.d, t.ta)$ is generated. Going back to Figure~\ref{fig:illustrative-rules}, a transition list $T$ fires via the [T-FIRE] rule when one of its constituent transitions fires. The [SD-FIRE] rule describes how a \stateflow{} state fires when one of the transitions from the $T_o$ transition list fires. According to the rule, the firing of the transition is followed by symbolic evaluation of the pending action from the existing $Fire$ event, then symbolic execution and evaluation of the actions and conditions of the currently active inner component(s), and finally the symbolic evaluation of the $sd.ex$ state action. The last two rules, [AND] and [OR-FIRE] capture the correct sequence of elements processing when executing the \texttt{And} and \texttt{Or-}compositions, respectively. Executing an \texttt{And-}composition involves processing each of its constituent states from the state definition list ($SD$), whereas an \texttt{Or-}composition fires when the underlying $sd$ fires. Due to space limitations, Figure~\ref{fig:illustrative-rules} shows only a small fraction of the rules for illustrative purposes. The complete set of 27 SSOS rules is included in Appendix~\ref{sec:appendix-A}. 

Since we are overloading the transition relation symbol ``$\mathit{\rightarrow}$'' in the SOS and SSOS rules, to avoid confusion, further in the paper we shall use ``$\mathit{\xrightarrow[]{SOS}}$'' for transitions derivable with the SOS rules, and ``$\mathit{\xrightarrow[]{SSOS}}$'' for transitions derivable with the SSOS rules.

\section{Characterization of the SSOS}
\label{sec:characterization-sos-ssos}

Our SSOS semantics is essentially an operational semantics for symbolic execution of \stateflow{} programs. It opens up the opportunity for application of a broader spectrum of verification techniques, such as: \emph{testing} (purely symbolic, or as a combination of symbolic and concrete (concolic) testing~\cite{godefroid05dartconcolictesting}) or \emph{bounded model checking}~\cite{biere2003boundedmodelchecking}. To be able to reason symbolically over \stateflow{} programs, however, one must first provide a formal characterization of the relationship between its concrete and symbolic execution. In this section, we prove two results that characterize this relationship. In Theorem~\ref{theo:theorem-1} we show that for each derivable SSOS transition there exists a corresponding derivable SOS transition. Conversely, in Theorem~\ref{theo:theorem-2} we show that for each derivable SOS transition there exists a derivable SSOS transition.
The connection is established in both cases by means of an interpretation of the symbolic values for which the Boolean expression added to the path condition holds. 

First, we introduce some additional notation.
Let $\beta: \mathit{SEnv} \times \mathit{Env} \rightarrow \mathit{Env}$ be a function that transforms a symbolic environment~$\Delta$ into a concrete one~$\beta (\Delta, D)$ with the help of an environment~$D$ that serves as an interpretation of the symbolic values; for any $v \in \emph{Var}$, let $\beta (\Delta, D) (v)$ be defined as the value of the expression~$\Delta (v)$ in the (renamed) environment $D \circ g^{-1}$. 
Similarly, let $\mathcal{B}: \mathit{BExpr}_{\mathit{Sym}} \rightarrow (\mathit{Env} \rightarrow \mathit{Bool})$ be a function that evaluates path conditions in concrete environments, so that $\mathcal{B} [\![\mathit{pc}]\!] (D)$ is the Boolean value of the path condition~$\mathit{pc}$ in $D \circ g^{-1}$.
Finally, observing that the transitions derived by the SSOS rules only (potentially) add a conjunct to the current path condition~$pc_k$ to obtain a new path condition~$pc^{k+1}$, let $pc^{k+1}_k$ denote this added conjunct (or~$\top$, if no conjunct is added). 

\begin{theorem}
\label{theo:theorem-1}
If $(P_1, \langle \Delta_1, pc_1\rangle) $ $\xrightarrow[]{SSOS}$ $(P_2, \langle \Delta_2, pc_2\rangle , tv)$, then for all $D_0 \in \mathit{Env}$ such that $\mathcal{B}[\![pc_1^2]\!](\beta(\Delta_1, D_0)) = \top$, we have $(P_1, \beta(\Delta_1, D_0))  \xrightarrow[]{SOS} (P_2, \beta(\Delta_2, D_0))$.
\end{theorem}

\begin{proof}
For proving the result, we shall use the principle of \textit{Rule Induction}~(see, e.g., \cite[p.~41]{winskel1993formalsemanticsofprogramminglanguages}). 
The principle states that, in order to prove that a given predicate over judgements holds for all judgements derivable by a given set of rules, one has to show that every rule preserves the predicate. In the statement of the theorem, the predicate to be proved is the one defined by the then-clause. 

The complete proof has thus to consider each of the 27 rules of the SSOS. We include here only a few selected cases of the proof to illustrate the proof technique. As it can be seen from the given cases, all of them follow the same pattern, so it should be clear to the reader how the complete proof unfolds. \\

\noindent \textbf{Case 1.1. [t-FIRE]\textsubscript{SSOS}}

Since the two premises of the rule are not transitions over which the statement is proved, we can treat them as side conditions and assume them to be true (since otherwise one cannot apply the rule). 
Next, let $\mathit{D_0 \in Env}$ be such that:
\begin{equation}
\label{eq:t-fire-main}
    \mathcal{B}[\![c]\!](\beta(\Delta_1, D_0)) = \top
\end{equation}
and let $D_1, D_2 \in Env \;.\; D_1 = \beta(\Delta_1, D_0)$ and $D_2 = $ $\beta(\Delta_2, D_0)$. Then, we have  $e \vdash (c, D_1) \rightarrow D_1$ and $\mathit{e \vdash (a, D_1) \rightarrow D_2}$, and therefore $e \vdash (t, D_1) \rightarrow D_2,$ $\mathit{Fire(d, ta)}$,
where $\mathit{pc_2 = pc_1 \land c}$ and $\mathit{\Delta_2 = \aexpr{a}(\Delta_1)}$. The premises of the [t-Fire]\textsubscript{SOS} rule are thus true, and we can apply the rule to obtain the following:

\begin{equation*}
    e \vdash (t, D_1) \rightarrow D_2, Fire(d, ta)
\end{equation*}

\noindent 
which is what needed to be demonstrated. 
\newline

\noindent \textbf{Case 1.2 [T-FIRE]\textsubscript{SSOS}}

Let $\mathit{D_0 \in Env}$, and $t = t.T$ be such that: 

\begin{equation}
    \label{eq:T-fire-main}
    \mathcal{B}[\![t.c]\!](\beta(\Delta_1, D_0)) = \top
\end{equation}
and let $\mathit{D_1, D_2 \in Env \;.\; D_1=}\beta\mathit{(\Delta_1, D_0), D_2=}\beta\mathit{(\Delta_2, D_0)}$.

\noindent From~(\ref{eq:T-fire-main}), and the induction hypothesis it follows: 
\begin{equation}
\label{eq:T-fire-premise}
    \mathit{e, J \vdash (t, D_1) \rightarrow D_2, Fire(d, ta)}
\end{equation}

Now that the premise for [T-FIRE]\textsubscript{SOS} rule given in~(\ref{eq:T-fire-premise}) is true, we can apply the rule to obtain the following:
\begin{equation*}
    \mathit{e, J \vdash (t.T, D_1) \rightarrow D_2, Fire(d, ta)}
\end{equation*}
\noindent which concludes the proof for the case.\newline

\noindent \textbf{Case 1.3 [SD-FIRE]\textsubscript{SSOS}}

Let $\mathit{D_0 \in Env}$ be such that:
\begin{equation}
    \label{eq:sd-int-fire-main}
    \begin{split}
    \mathcal{B}[\![pc_1^2]\!](\beta(\Delta_1,& D_0)) = \top, \; \mathcal{B}[\![pc_2^3]\!](\beta(\Delta_3, D_0)) = \top, \; \mathcal{B}[\![pc_3^4]\!](\beta(\Delta_4, D_0)) = \top
    \end{split}
\end{equation}
\noindent where $\mathit{pc_2 = pc_1 \land pc_1^2,\: pc_3 = pc_2 \land pc_2^3,\: pc_4 = pc_3 \land pc_3^4}$.  
Based on~(\ref{eq:sd-int-fire-main}), we know that 
\begin{equation*}
    \begin{split}
        \exists D_1, \dots, D_5 \in Env \;.\; &D_1 =\beta(\Delta_1, D_0), D_2 =\beta(\Delta_2, D_0), D_3 = \beta(\Delta_3, D_0), \\
         &D_4 =\beta(\Delta_4, D_0), D_5 =\beta(\Delta_5, D_0)
    \end{split}
\end{equation*}
\noindent From~(\ref{eq:sd-int-fire-main}) and the induction hypothesis, it follows that: 
\begin{equation}
\label{eq:sd-fire-premises}
    \begin{split}
    &e, J_0 \vdash (T_0, D_1) \rightarrow D_2, Fire(d, ta); \; e \vdash (a, D_2) \rightarrow D_3; \\&
    \; e, J \vdash (C, D_3) \rightarrow D_4;\; e, J \vdash (A.ex, D_4) \rightarrow D_5
    \end{split}
\end{equation}
Since the premises for [SD-FIRE]\textsubscript{SOS} rule given in~(\ref{eq:sd-fire-premises}) are true, we can apply the rule to obtain the following:
\begin{equation*}
        e, J_O \vdash ((A,C,T_i,T_o,J), D_1) \rightarrow{}  ((A,C',T_i,T_o,J), D_5), Fire(d,\diamond)
\end{equation*}
\noindent which is what needed to be shown. 
\newline 

\noindent \textbf{Case 1.4 [AND]\textsubscript{SSOS}}
Let $D_0 \in Env$ be such that the following holds:
\begin{equation}
    \label{eq:and-main}
    \forall i \in [1, \dots, n] \;.\; \mathcal{B}[\![pc_i^{i+1}]\!](\beta(\Delta_i, D_0)) = \top;
\end{equation}
From the initial assumption and~(\ref{eq:and-main}), we know that:
\begin{equation*}
    \mathit{\forall k \in [1,\; \dots,\; n+1] \;.\; \exists D_k \in Env \;.\; D_k = \,} \beta\mathit{(\Delta_k, D_0)}
\end{equation*}
From the induction hypothesis and~(\ref{eq:and-main}), it follows that: 
\begin{equation}
\label{eq:and-sos-premise}
    \mathit{\forall i \in [1, \dots, n] \quad e, J \vdash (sd_i, D_i) \rightarrow sd_{i}', D_{i+1}, No}
\end{equation}
\noindent Since the premise for the [AND]\textsubscript{SSOS} rule~(\ref{eq:and-sos-premise}) is true, we can apply the rule to obtain the following:
\begin{equation*}
    \begin{split}
        e,J \vdash &(e,J \vdash (And\{s_0:sd_0 \cdots s_n:sd_n\}, D_1) \rightarrow{}\\ 
        &(And\{s_0:sd'_0 \cdots s_n:sd'_n\}, D_{n+1}, No)
    \end{split}
\end{equation*}

\noindent which is what was needed to be shown. 
\newline 

\noindent \textbf{Case 1.5 [OR-FIRE]\textsubscript{SSOS}}

Let $D_0 \in Env$ be such that the following holds:

\begin{equation}
        \label{eq:or-fire-main}
    \mathcal{B}[\![pc_1^{2}]\!](\beta(\Delta_1, D_0)) = \top;
\end{equation}
Based on~(\ref{eq:or-fire-main}), the following holds:

\begin{equation*}
    \mathit{\exists D_1, D_2 \in Env \;.\; D_1 = }\beta\mathit{(\Delta_1, D_0), D_2 = }\beta\mathit{(\Delta_2, D_0)}
\end{equation*}
From the inductive hypothesis and~(\ref{eq:or-fire-main}), it follows that: 
\begin{equation}
\label{eq:or-fire-premise}
\mathit{e,J \vdash (sd, D_1) \rightarrow (sd', D_2), Fire(p',a)}
\end{equation}
\noindent Since the premise for the [OR-FIRE]\textsubscript{SOS} rule~(\ref{eq:or-fire-premise}) is true, we can apply the rule to obtain the following:
\begin{equation*}
 \begin{split}
    e,J \vdash &(Or(s,p,T,SD[s:sd]), D_1, tv)\rightarrow{}(Or(\emptyset_s,p,T,SD[s:sd']), D_2, Fire(p',a))
     \end{split}
\end{equation*}
\noindent which is what was needed to be shown.
\end{proof}

Our next result establishes the reverse direction.

\begin{theorem}
\label{theo:theorem-2}
If $(P_1, D_1) \xrightarrow[]{SOS} (P_2, D_2)$, then for all $pc_1 \in \mathit{BExpr_{Sym}}$, $\Delta_1 \in \mathit{SEnv}$ and $D_0 \in \mathit{Env}$ such that $\beta(\Delta_1, D_0) = D_1$, there exist $pc_2,\; pc_1^2 \in \mathit{BExpr_{Sym}}$ and $\Delta_2 \in \mathit{SEnv}$ such that $pc_2 = pc_1 \land pc_1^2$, $\mathcal{B}[\![pc_1^2]\!](\beta(\Delta_1, D_0)) = \top$, $\beta(\Delta_2, D_0) = D_2$ and $(P_1, \langle\Delta_1, pc_1 \rangle) \xrightarrow[]{SSOS} (P_2,  \langle\Delta_2, pc_2 \rangle)$.
\end{theorem}

\begin{proof}
Again, the proof is by Rule Induction. We only show two cases here. The rest of the cases are proved by following the scheme of the presented ones. \\

\noindent \textbf{Case 2.1. [t-FIRE]\textsubscript{SOS}}

Let us assume that a \stateflow{} program performs a concrete transition derivable using the [t-Fire]\textsubscript{SOS} rule. Lets assume that for arbitrary $\mathit{\langle \Delta_1, pc_1\rangle \in SS}$ the following holds:
\begin{equation}
\label{eq:t-fire-main-sos}
    \exists D_0 \in Env \: . \: \beta(\Delta_1, D_0) = D_1
\end{equation}
Also, we know that $\mathit{pc_1^2 = \mathcal{B}[\![c]\!](\beta(\Delta_1, D_0)) = \top}$, due to the inductive hypothesis. Consequently, the following also holds: $\mathit{\mathcal{B}[\![pc_1 \land c]\!](\beta(\Delta_1, D_0)) = \top}$. Based on~(\ref{eq:t-fire-main-sos}) and the true valuation of $\mathit{pc_1 \land c}$, the following also holds:
\begin{equation*}
    e \vdash (t, \beta(\Delta_1, D_0)) \rightarrow \beta(\Delta_2, D_0), Fire(d, ta)
\end{equation*}
where $\mathit{pc_2 = pc_1 \land c}$ and $\mathit{\Delta_2 = \aexpr{a}(\Delta_1)}$. Based on~(\ref{eq:t-fire-main-sos}) and the induction hypothesis we derive the following: $\mathit{e \vdash (c, \langle \Delta_1, pc_1\rangle) \rightarrow \langle \Delta_1, pc_2\rangle}$ and $\mathit{e \vdash (a, \langle \Delta_1, pc_2\rangle) \rightarrow \langle \Delta_2, pc_2\rangle}$. If we now apply the [t-Fire]\textsubscript{SSOS} rule over the last two premises, we derive the following transition:
\begin{equation*}
e \vdash (t, \langle\Delta_1, pc_1\rangle) \rightarrow \langle\Delta_2, pc_2\rangle, Fire(d, ta))
\end{equation*}

\noindent which is what needed to be demonstrated. 
\newline

\noindent \textbf{Case 2.2 [T-FIRE]\textsubscript{SOS}}

Let us assume that a \stateflow{} program performs a concrete transition derivable using the [T-Fire]\textsubscript{SOS} rule. Let us assume that for an arbitrary $\langle \Delta_1, pc_1\rangle \in SS$ the following holds:
\begin{equation}
    \label{eq:T-fire-main-sos}
     \exists D_0 \in Env \: . \: \beta(\Delta_1, D_0) = D_1
\end{equation}
Also, we know that $pc_1^2 = \mathcal{B}[\![c]\!](\beta(\Delta_1, D_0)) = \top$, due to the inductive hypothesis. Consequently, the following also holds: $\mathcal{B}[\![pc_1 \land c]\!](\beta(\Delta_1, D_0)) = \top$. Based on~(\ref{eq:T-fire-main-sos}) and the truth valuation of the $pc_1 \land c$, the following also holds:
\begin{equation*}
    e \vdash (t, \beta(\Delta_1, D_0)) \rightarrow \beta(\Delta_2, D_0), Fire(d, ta)
\end{equation*}
where $pc_2 = pc_1 \land c$ and $\Delta_2 = \aexpr{a}(\Delta_1)$. Based on~(\ref{eq:T-fire-main-sos}) and the induction hypothesis we derive the following: $e \vdash (c, \langle \Delta_1, pc_1\rangle) \rightarrow \langle \Delta_1, pc_2\rangle$ and $e \vdash (a, \langle \Delta_1, pc_2\rangle) \rightarrow \langle \Delta_2, pc_2\rangle$. If we now apply the [T-Fire]\textsubscript{SSOS} rule over the last two premises, we derive the following transition:
\begin{equation*}
e \vdash (t, \langle\Delta_1, pc_1\rangle) \rightarrow \langle\Delta_2, pc_2\rangle, Fire(d, ta)
\end{equation*}
\noindent which is what needed to be demonstrated. 
\end{proof}

There are two important corollaries of the above two results, which we will only state here informally. First, both results lift naturally to \emph{executions}, i.e., to sequences of transitions. Note in particular how in Theorem~\ref{theo:theorem-2} the ``for all~$\mathit{pc}_1$ \ldots there exists~$\mathit{pc}_2$'' part allows the sequential composition of transitions. Second, when starting from a true path condition, as one does in symbolic execution, the \emph{satisfying assignments} for the path condition at the end of any symbolic path, viewed as interpreting environments, define precisely the concrete paths that follow the symbolic one. 

Further, the executions in SOS and SSOS can be shown to \emph{simulate} each other with respect to processing external events. It is well-known that invariant properties are preserved by simulation, and thus, can be checked by symbolically executing the given \stateflow{} program. 
Even if limited, this class of properties is important in industrial contexts, as our collaboration with Scania on formally verifying safety-critical embedded code generated from Simulink models has shown.

\section{From \stateflow{} Programs to SMT Solving}
\label{sec:transformation-and-smt-encoding}

In our work, we focus on using BMC for checking invariant properties over symbolic representation of \stateflow{} programs. In Section~\ref{sec:symbolicsos} we developed an SSOS for \stateflow{}, and exhibited in Section~\ref{sec:characterization-sos-ssos} a simulation relation between executions derived in SOS and SSOS, which is sufficient for the preservation of invariant properties. In the following, we show how we use the SSOS to relate \stateflow{} programs to STS over symbolic configurations. We define the \emph{k-bounded invariant checking} problem for the latter representation (Section~\ref{sec:stateflow-to-sts}), and show how this problem can be encoded as an SMT problem (Section~\ref{sec:sts-to-smt}).

\subsection{Bounded Invariant Checking for Stateflow Programs}
\label{sec:stateflow-to-sts}

In this section, we define a version of STS that encode the \emph{symbolic} behaviors of \stateflow{} programs, and then adapt the BMC problem to such transition systems.

\begin{definition}[STS over Symbolic Configurations]
\label{def:sts-kappa}
A \emph{symbolic transition system over the symbolic configurations} of a given \stateflow{} program is an STS $\widehat{S} = (\widehat{I}, \widehat{R})$, in the sense of Definition~\ref{def:sts-with-predicates}, but over the symbolic configurations and transitions of the program as induced by the SSOS rules. 
\end{definition}
\vspace{-2mm}
\noindent $\widehat{I}(\cdot)$ and $\widehat{R}(\cdot, \cdot)$ are thus a unary ``initialization'' predicate and a binary ``next-state" predicate over the symbolic configurations of the program, respectively, which are quantifier-free FOL formulas over the components of symbolic configurations. 

The formal relationship between an STS over symbolic configurations~$\widehat{S}$ and an ordinary STS~$S$ of a \stateflow{} program is given by the following result. %

\begin{proposition}
\label{theo:sts-sts-kappa-relationship}
Let SF be a \stateflow{} program, $S = (I, R)$ be an STS over its concrete configurations as induced by the SOS rules, and $\widehat{S} = (\widehat{I}, \widehat{R})$ be an STS over its symbolic configurations as induced by the SSOS rules. Then, the following equivalences hold: \\

\begin{equation*}
    (1)~\widehat{I}(P, \langle pc, \Delta \rangle) \>\Leftrightarrow\> \, \exists D_0 \in \mathit{Env}.\  I(P, D_0) \>\land\> I(P, \beta(\Delta, D_0)) \>\land\> \mathcal{B}[\![pc]\!](\beta(\Delta, D_0))
\end{equation*}

\begin{equation*}
    \begin{split}
        (2)~ \widehat{R}((P, \langle pc, \Delta \rangle), (P', \langle pc', \Delta' \rangle)) & \Leftrightarrow\> \\
        & \exists D_0 \in \mathit{Env}. \mathcal{B}[\![pc]\!](\beta(\Delta, D_0)) \land
  \mathcal{B}[\![pc']\!](\beta(\Delta', D_0)) \>\\
        &\land \ R((P, \beta(\Delta, D_0)), (P', \beta(\Delta', D_0)))) 
    \end{split}
\end{equation*}
\end{proposition}

\begin{proof}
~\\\noindent\textbf{\textit{Direction~$(\Rightarrow)$ of (1)}.}\\
Assume that $\widehat{I}(P, \langle\Delta, pc\rangle) = \top$. From the definition of initial symbolic configuration (Section~\ref{sec:symbolicsos}), we know that that $pc = pc_0$ and $\Delta = \Delta_0$, and that $pc_0 = \top$ and $\Delta = g$. Since $pc = \top$, there must exist $D_0 \in Env$ such that $I(P, D_0)$, and $\beta(\Delta, D_0) = D$. Given that $\Delta = g$, it follows that $D = D_0$. Since $I(P, D_0) = \top$, then also $I(P, D) = \top$.  \\

\noindent\textbf{\textit{Direction~$(\Leftarrow)$ of (1}).}\\
Assume that there exists $D_0 \in Env$ such that $I(P, D_0)$, $\mathcal{B}[\![pc]\!](\beta(\Delta, D_0)) = \top$, and $I(P, \beta(\Delta, D_0))$. If $I(P, \beta(\Delta, D_0)) = \top$, then $\beta(\Delta, D_0) = D$. $I(P, D_0) = \top$ and $I(P, D) = \top$ if and only if $D = D_0$. This is possible only if $\Delta = g$. Given the definition of initial symbolic configuration (Section~\ref{sec:symbolicsos}), the assumption $\mathcal{B}[\![pc]\!](\beta(\Delta, D_0)) = \top$, and $\Delta = g$, it follows that $\widehat{I}(P, \langle\Delta, pc\rangle) = \top$. \\

\noindent\textbf{\textit{Direction~$(\Rightarrow)$ of (2).}}\\
Assume that $\widehat{R}((P, \langle\Delta_1, pc_1\rangle), (P', \langle \Delta_2, pc_2 \rangle)) = \top$. According to Definition 3, if $\widehat{R}((P, \langle \Delta_1, pc_1 \rangle), (P', \langle \Delta_2, pc_2 \rangle)) = \top$, then there exists $(P, \langle \Delta_1, pc_1 \rangle) \xrightarrow{SSOS} (P', \langle \Delta_2, pc_2 \rangle)$. From Theorem~1, if $(P, \langle \Delta_1, pc_1 \rangle) \xrightarrow{SSOS} (P', \langle \Delta_2, pc_2 \rangle)$, then for all $D_0 \in Env$ such that $\mathcal{B}[\![pc_1]\!](\beta(\Delta_1, D_0)) = \top$, and $\mathcal{B}[\![pc_1^2]\!](\beta(\Delta_1, D_0)) = \top$, there is $(P, \beta(\Delta_1, D_0)) \xrightarrow{SOS} (P, \beta(\Delta_2, D_0))$. Using the result for the existence of an SOS transition, and Definition~2, we can conclude that:
\begin{equation*}
    R((P, \beta(\Delta_1, D_0)), (P, \beta(\Delta_2, D_0))) = \top    
\end{equation*}

\noindent\textbf{\textit{Direction~$(\Leftarrow)$ of (2).}}\\
Assume that there exists $D_0 \in Env$ such that $\mathcal{B}[\![pc_1]\!](\beta(\Delta_1, D_0)) = \top$, $\mathcal{B}[\![pc_2]\!](\beta(\Delta_2, D_0)) = \top$, and $R((P, \beta(\Delta_1, D_0)), (P', \beta(\Delta_2, D_0))) = \top$. From Definition~2, if $R((P, \beta(\Delta_1, D_0)), (P, \beta(\Delta_2, D_0))) = \top$, then $(P, \beta(\Delta_1, D_0))$ $\xrightarrow{SOS} (P, \beta(\Delta_2, D_0))$. 

Based on the existence of $(P, \beta(\Delta_1, D_0)) \xrightarrow{SOS} (P, \beta(\Delta_2, D_0))$, and by following Theorem~2, we have that for all $pc_1 \in BExpr_{Sym}$, $\Delta_1 \in SEnv$, and $D_0 \in Env$ such that $\beta(\Delta_1, D_0) = D_1$, there exist $pc_1^2, pc_2 \in BExpr_{Sym}$, and $\Delta_2 \in SEnv$ such that $pc_2 = pc_1 \land pc_1^2$, $\mathcal{B}[\![pc_1^2]\!](\beta(\Delta_1, D_0)) = \top$, $\beta(\Delta_2, D_0) = D_2$ and $(P, \langle \Delta_1, pc_1\rangle) \xrightarrow{SSOS} (P', \langle\Delta_2, pc_2\rangle)$. Using the result for the derived SSOS transition, and Definition~3, we can conclude that $\widehat{R}((P, \langle\Delta_1, pc_1\rangle), (P', \langle\Delta_2, pc_2\rangle)) = \top$. 
\end{proof}

Now, let~$\varphi$ be a predicate over the concrete configurations of a \stateflow{} program. Predicate $\varphi$ induces a corresponding predicate $\widehat{\varphi}(sc) \triangleq \varphi(sc[g^{-1}])$ over the symbolic configurations $\mathit{sc = (P, \langle pc, \Delta \rangle)}$, where~$g$ is the bijection from Section~\ref{sec:symbolicsos}. Assuming an interpretation for the \emph{path} and \emph{k-bounded invariant property} formulas for executions over symbolic configurations, the counter-example path formula~(\ref{eq:counter-example-definition}) for symbolic executions can be rewritten as follows:

\begin{equation}
\label{eq:counter-example-definition-sts-kappa}
    \exists sc_0, \dots, sc_k.\ (path(sc_0, \dots, sc_k) \land 
    \bigvee\limits_{i=0}^{k} \neg\widehat{\varphi}(sc_i))
\end{equation}~

\noindent
Based on formula~(\ref{eq:counter-example-definition-sts-kappa}), and along the lines of the definition of BMC for C programs given in~\cite{armando2009smt-bmc}, we derive the following.
\begin{theorem}
\label{theo:sts-kappa-bmc}
Let SF be a \stateflow{} program, $\widehat{S} = (\widehat{I}, \widehat{R})$ be an STS over its symbolic configurations, and $\varphi^k$ be a k-bounded invariant property. Then, the following two statements are equivalent:

\begin{enumerate}
 \item SF satisfies the k-bounded invariant property $\varphi^k$.
 \item The formula $path(sc_0, \dots sc_k) \land \bigvee\limits_{i=0}^{k} \neg\widehat{\varphi}(sc_i)$ is UNSAT. 
\end{enumerate}
\end{theorem}

\begin{proof} (By contradiction.)
Assume that a given \stateflow{} program does not satisfy the \emph{k-}bounded invariant property~$\varphi^k$, and that statement (2) holds. From the definition of \emph{k}-bounded invariant property, we know that such a property fails if there exists a path in which the last configuration violates~$\varphi$. From Definition~\ref{eq:counter-example-definition-sts-kappa}, we know that such a path exists if the formula given in (2) is satisfiable (SAT), which contradicts the initial assumption.

Assume that the formula in (2) is SAT, and that the statement (1) holds. The satisfiability of the formula says that there exists a reachable state $sc_i. \; i < k$, in which the negation of the property $\widehat{\varphi}$ holds, that is, $\widehat{\varphi}(sc_i) = \top$. In such case $\widehat{\varphi}^k$ is not an invariant, which contradicts our initial assumption.
\end{proof}

Now that we have formally defined BMC invariant checking for STS over symbolic configurations, we show how to construct the predicates~$\widehat{I}$ and~$\widehat{R}$ for a given \stateflow{} program. 

\subsection{From \stateflow{} Programs to SMT Scripts}
\label{sec:sts-to-smt}

In this section, we describe a procedure for deriving an STS from a given \stateflow{} program using the set of SSOS rules, and the transformation of the STS predicates into quantifier-free FOL formulas that can be used for \emph{k-bounded invariant checking} over symbolic configurations, as defined in Theorem~\ref{theo:sts-kappa-bmc}.

Before deriving an STS $\widehat{S} = (\widehat{I}, \widehat{R})$ from a given \stateflow{} program, one must make a decision as to the \emph{granularity} of the transitions in executions. At its highest syntactic level, any \stateflow{} program is an \texttt{Or}-composition or an \texttt{And}-composition. Since our running example is an \texttt{Or}-composition, and since conceptually the procedure for \texttt{And}-compositions is virtually the same, in the following we focus on the \texttt{Or}-compositions only. We derive an STS in which the transitions between configurations correspond to transitions at the top \texttt{Or}-composition level. Due to the layered structure of the imperative language, each such transition consists of a series of transitions at the lower levels (see Table~\ref{tab:imperative-language} from Section~\ref{sec:h-r-sos}). Our approach to the derivation of the top-level transitions is to use our SSOS to perform \emph{symbolic execution} between any possible pair of consecutive control points of the program, for arbitrary values of the data. One should note that in general case, the derivation of the STS is not strictly bound to the top-level component, as it can be done against any syntactic class of the Stateflow imperative language.

As a result of our adopted modeling principle, the configurations for the induced STS are of the following type: $(Or, \langle pc, \Delta\rangle)$. Even though the program component during execution remains the same (the top-level \texttt{Or}-component), it can be the case that its internal configuration changes. The internal configuration of an \texttt{Or}-component is characterized by the set of active substates. Consequently, the \emph{program control points} correspond to the possible internal configurations at the top \texttt{Or}-composition level.
For instance, our running \stateflow{} example from Figure~\ref{fig:stateflow-scenario} has 5 program control points, namely: i) no active states, ii) \emph{Stop} and \emph{Reset}, iii) \emph{Stop} and \emph{Lap\_stop}, iv) \emph{Run} and \emph{Running}, and v) \emph{Run} and \emph{Lap}. 
One way of modeling the program control points as defined above is to extend the set of variables $\mathit{Var}$ with a set of auxiliary Boolean variables ($\mathit{Var_C}$). For every control point~$\mathit{Or}$, the set of variables $\mathit{Var_C}$ can be partitioned into two subsets: the set $\mathit{Var_{C^+} = \{v \: | \: v \in \mathit{Var_C}, \: v = \top\}}$ corresponding to the active states of~$\mathit{Or}$, and $\mathit{Var_{C^-}} = \mathit{Var_C} \setminus \mathit{Var_{C^+}}$. Thus, the control point~$\mathit{Or}$ is characterized by the formula:
\begin{equation}
\label{eq:orphi}
    \Phi_{\mathit{Or}} \>\triangleq\> \bigwedge\limits_{v \, \in \, Var_{C^+}} v \; \land \bigwedge\limits_{v \, \in \, Var_{C^-}} \neg v
\end{equation}

The initial symbolic configuration for an STS of a \stateflow{} program given as an \texttt{Or}-composition is defined as $(Or_\emptyset, \langle pc_0, \Delta_0\rangle)$, where: $Or_\emptyset$ is an \textit{uninitialized}  \texttt{Or}-component that has no active substates, $pc_0 = \top$, and $\Delta = g$ (see Section~\ref{sec:symbolicsos}). As explained, in our STS we only have transitions between symbolic configurations of the form $\mathit{(Or, \langle pc, \Delta \rangle)} \rightarrow \mathit{(Or', \langle pc', \Delta' \rangle)}$, where $Or$ and $Or'$ are consecutive program control points. To express the predicates $\widehat{I}$ and~$\widehat{R}$ as quantifier-free FOL formulas, we need to construct the quantifier-free FOL formulas for the~$pc$ and~$\Delta$ components of the symbolic configurations.

The path condition~$pc$ is a quantifier-free Boolean expression over symbols, and as such can be viewed as a quantifier-free FOL formula $\Phi_{pc}$. The symbolic environment~$\Delta$, on the other hand, is a mapping between program variables and arithmetic expressions over symbols. From~$\Delta$, one can construct a quantifier-free FOL formula modulo theory of arithmetic for~$\Phi_\Delta$, as follows: 

\begin{equation}
\label{eq:deltaphi}
\begin{aligned}
    \Phi_\Delta \triangleq \bigwedge\limits_{v \in \mathit{Var}_D} v' = \Delta(v)
\end{aligned}
\end{equation}

\noindent where $\mathit{Var_D} = \mathit{Var} \setminus \mathit{Var_C}$.

Now that we have defined the construction of quantifier-free FOL formulas for each of the components of the symbolic configurations of an STS, we can construct, for any transition~$T_i$ between symbolic configurations, a quantifier-free FOL formula~($\Phi_{T_i}$) modulo theory of arithmetic, as follows:

\begin{equation}
\label{eq:transitionphi}
    \Phi_{T_i} \>\triangleq\> \Phi_{Or} \land \Phi_{pc_1^2} \Rightarrow \Phi_{Or'} \land \Phi_{\Delta'}
\end{equation}~

Finally, based on the formula~(\ref{eq:transitionphi}) and Proposition 1, we encode the predicates $\widehat{I}$~and~$\widehat{R}$ as the following quantifier-free FOL modulo theory of arithmetic formulas:

\begin{equation}
\label{eq:i-and-r-fol}
    \begin{aligned}
        &\widehat{I} \ \triangleq \ \Phi_{\mathit{Or}_\emptyset} \land \Phi_{\Delta_0} \\
        &\widehat{R} \ \triangleq \ \bigwedge\limits_{T_i \in T} \Phi_{T_i}
    \end{aligned}
\end{equation}
\noindent where $T$ is the set of all derivable SSOS transitions from the initial top-level composition. One way of computing~$T$ is to start from the initial program control point ($Or_\emptyset$) and derive all transitions between reachable program control points with a standard breadth-first or depth-first search algorithm.

The remaining step now is to encode the FOL formulas corresponding to the $\widehat{I}$ and $\widehat{R}$ predicates into the SMT-LIB script format, in order to be used as an input to an SMT-solver. We assume that the program data ($\mathit{Var_D}$) in the \stateflow{} program is composed of scalar variables, which are either numeric or Boolean. There are two ways of encoding the numeric variables for SMT: i)~using the theory of bit-vectors~\cite{ganesh07bitvector}, or ii)~directly as variables of the sort that corresponds to their numeric domain. The bit-vector encoding provides an accurate way of capturing the binary representation of the numeric values; however, the resulting formula depends on the size of the vector. The numeric representation, on the other hand, provides an encoding independent of the binary representation, but does not guarantee completeness for non-linear arithmetic expressions. In our case, the latter suffices. We model the \emph{events} using Boolean variables.

An earlier work shows how a sequence of C~commands can be converted into an SMT script~\cite{armando2009smt-bmc}, using the concepts of \emph{conditional normal form}, and \emph{single static assignment form} (SSA) (see, e.g., \cite{cytron1989ssa}). A program is in conditional normal form if all the statements are of the form \emph{if (c) then r}, where \emph{r} is either an assertion or an assignment. The SSA encoding principle requires each variable to be assigned only once, which means that for each assignment a \emph{fresh} variable is added to the set $\mathit{Var}$. For instance, $v_1^1$ is a fresh variable that represents the first intermediate assignment of the variable $v$ during the second execution step. This resolves the SMT-encoding for the sequential execution of the accumulated actions from the derived transitions.

\section{Practical Evaluation of \sesf{}}
\label{sec:evaluation-and-comparison}

In this section, we present the preliminary results from the initial practical evaluation of our approach (henceforth referred to as \sesf{}). As benchmark, we use the Stopwatch running example from Section~\ref{sec:stateflow}. The main purpose of this evaluation is to obtain preliminary data for the practical usefulness of our approach, in terms of its ability to detect design errors, and the time it requires to detect them. Even though the most natural way to assess the applicability and the practical usefulness of our approach is to benchmark it against the SLDV tool on a wider set of use cases, in the end it was not possible due to the licensing constraints described in Section~\ref{sec:introduction}.

To this date, we have automated the following aspects of our approach: the generation of an underlying STS $\widehat{S} = (\widehat{I}, \widehat{R})$, the incremental unrolling of the predicate $\widehat{R}$ that models the program execution, and checking the satisfiability of the generated formula after each unrolling, using the Z3 solver. The only part that is not yet automated is the SMT-LIB encoding of the underlying STS.

Before we delve into the analysis, we explain the environmental setup for the evaluation. Our \sesf{} tool requires that the \stateflow{} model is encoded into the imperative language, which is then provided as an input to the \sesf{} tool. As the transformation of the graphical model into an imperative program is beyond the scope of this paper, we assume that there exists a tool that converts an original \stateflow{} model into a program written in the imperative language. The complete evaluation is performed on a standard MacBook Pro workstation with a 2,4 GHz Quad-Core Intel Core i5 processor and 8GB of RAM memory. We use Matlab version R2020b.

The set~$T$ of symbolic transitions at the program level is generated by a \emph{symbolic execution engine} which is implemented in Python. For each of the transitions, the symbolic execution engine generates a certificate for the correctness of the transition in the form of an SSOS \emph{derivation tree}~\cite{sesfGitLink}. The tool generates the set $T$ for the Stopwatch program, consisting of 22~transitions, within one second. Next, we manually encode the set~$T$ into SMT-LIB assertions, by following the procedure described in Section~\ref{sec:sts-to-smt}. 

\begin{figure}[t!]
\centering
\resizebox{.7\columnwidth}{!}{
\begin{tikzpicture}
\begin{axis}[
    title={},
    xlabel={Length of the counter-example path},
    ylabel={Required time (in seconds)},
    xmin=0, xmax=100,
    ymin=0, ymax=50,
    xtick={0,10,20,30,40,50,60,70,80,90,100},
    ytick={10,20,30,40,50},
    legend pos=north west,
    ymajorgrids=true,
    xmajorgrids=true,
    grid style=dashed,
    legend cell align={left},
]
\addplot[
    color=blue,
    mark=square,
    ]
    coordinates {
    (5,0.909)(10,1.026)(20,1.567)(40,3.672)(80,13.602)(99,25.963)
    };
    \end{axis}
\end{tikzpicture}
}
\caption{Evaluation results} \label{fig:fig:evaluation-results}
\end{figure}
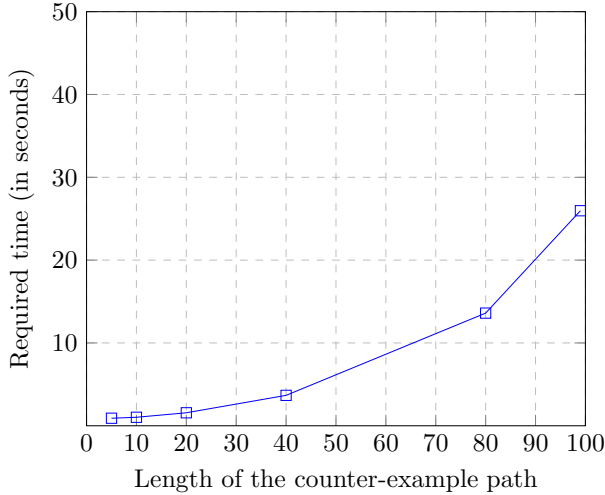

Next, we proceed to the main part of the evaluation, where we measure the time required by the tool to find a violation for a given invariant property, i.e., a counter-example, which is completely automated. In order to test the time for finding a counter-example, we analyze the Stopwatch model against the following parametric invariant property: \emph{The value of cent is always between 0 and X}, for which we know that there exists a counter-example trace of a specific length since the $cent$ variables will increase up to 100. We then run the analysis for the following set of values for \emph{X}: $\{5, 10, 20, 40, 80, 99\}$. The \sesf{} was able to provide a counter-example for each instance of the property. The results, in terms of time required to find the counter-example for each of the values of~\emph{X}, are given in Figure~\ref{fig:fig:evaluation-results}. 

Following the data plot, we can see that the \sesf{} tool was able to find violation for all the parameter values for the parametric safety property in a time frame ranging from several seconds to less than two minutes. Even though the obtained data can be best judged in a head-to-head comparison with the state-of-the-practice tool for formal verification of \stateflow{} models, we are unable to do so due to the aforementioned licensing issues. In spite of the limited validation presented in this section, we deem our approach and the prototype implementation to have potential for practical application as it was able to find the injected design errors within a reasonable time frame.

\section{Related Work}
\label{sec:related-work}

To the best of our knowledge, the present work is the first to define a symbolic operational semantics for any kind of \stateflow{} language, and a corresponding notion of STS over symbolic configurations. The main goal of this paper has been to establish the foundations for provably correct verification of \stateflow{} models using symbolic methods, while the development of a practical tool and its extensive evaluation has mostly been left as future work. Still, in this section we will outline some of the state-of-the-art approaches and tools for the formal analysis and verification of \stateflow{} models, and will draw some high-level parallels to our work.

The most common way of formally verifying \stateflow{} models is to propose and apply transformation rules for the basic \stateflow{} modeling constructs into some existing formal framework. Yang et al.~\cite{yang16stateflowtransformationase,jiang2019dependablecpsstateflow} propose a technique for transforming \stateflow{} models into \uppaal{} timed automata. The main idea of the technique is that by translating a \stateflow{} model into a timed automaton one creates a timed model, which can then be subjected to verification with respect to \emph{timing} and \emph{liveness} properties, based on an informal semantics of the \stateflow{} language. A similar endeavour is by Chen~\cite{chen10formalanalysisofstateflowdiagrams}, where an approach for formal analysis of \stateflow{} diagrams based on the PAL model checker is presented. Although claimed by the author that all of the \stateflow{} modeling principles are covered, the correctness of the formal model cannot be formally demonstrated. Meenakshi et al.~\cite{meenakshi2006toolfortranslatingsimulink} propose a tool for the transformation of Simulink models into the input language of a model checker. Similarly to the previous work, the main limitation of this work is proving the correctness of the transformation. Miyazawa et al.~\cite{miyazawa12refinementorientedmodelsofstateflow} provide a formalization of \stateflow{} in a refinement language called Circus. The main difference with our work lies in the fact that the they define formal semantics specific for the Circus language, whereas in our case the semantics are defined in generalized SOS-style. The consequence of this is the usability of the semantics, which in the case of Miyazawa et al.~\cite{miyazawa12refinementorientedmodelsofstateflow} are tool-specific, whereas in our case the proposed SSOS are not particularly bounded to any verification method or tool.  

There exists a set of approaches that treat \stateflow{} models as hybrid system models, and apply corresponding modelling and analysis approaches. In this category, we can mention the approach by Alur et al.~\cite{alur2008symbolicanalysisforcoveragesimulinkstateflow}, which uses a combination of numerical simulation and symbolic analysis for Simulink/\stateflow{} models for improving the simulation coverage of the models. At the core of their technique lies a transformation of Simulink/\stateflow{} models into \emph{linear hybrid systems}, which are then used for the analysis of the models using backward image computation, in order to identify classes of inputs, which can then be analyzed using minimal numerical simulations. Zuliani et al.~\cite{zuliani10bayesiansimulinkstateflow} employ statistical model checking (SMC) based on Bayesian statistics to reason about the correctness of \stateflow{} models. By resorting to SMC they alleviate the state-space explosion problem of complex \stateflow{} models; however, the obtained results are in a form of probabilistic bounds, which cannot be directly applied for refining the \stateflow{} models. Duggirala et al.~\cite{duggirala15c2e2} propose the C2E2 tool, which can be used for verification of a broad class of hybrid and dynamic system models based on validated simulations. In contrast, the aim of our work is to establish the foundations for a symbolic analysis of completely discrete-time \stateflow{} models.

There is a body of work that is focused on defining a denotational semantics for \stateflow{} models. Hamon~\cite{hamon2005denotational} proposed a denotational semantics for the \stateflow{} imperative language, as a complement to the already existing SOS semantics, in order to formalize the compilation of \stateflow{} programs. The denotational semantics is particularly useful for code generation from \stateflow{} models. In later work, Bourbouh et al.~\cite{bourbouh2017automatedanalysisofstateflowmodels,bourbouh2020cocosim} extended and upgraded the denotational semantics of Hamon, and employed it for the compilation of \stateflow{} models into hierarchical state machines, and later also into Lustre models, for SMT-based formal analysis and verification, as well as for code generation. As this approach is very similar to ours, we aim to perform in the future a more detailed comparison, especially in terms of the generated analysis model as it is generated based on a different type of semantics. Auto-Gene~\cite{toom06geneauto} is an open-source toolset intended for code generation from high level modelling languages, including Simulink and \stateflow{}. The aim of the toolset is to use among others Simulink and \stateflow{} as input languages, from which it will automatically generate C code. It uses only a safe subset of the Simulink/\stateflow{} language in order to be able to provide support for the verification of properties such as termination, context independence, etc.

\section{Conclusion}
\label{sec:conclusions}

We presented a technique for symbolic analysis of \stateflow{} programs with respect to invariant properties using bounded model checking (BMC). To this end, we developed a symbolic structural operational semantics (SSOS) for the \stateflow{} language. Our semantics is built on top of the work by Hamon and Rushby~\cite{hamon2004operationalsemanticsstateflow,hamon2007operationalsemanticsstateflow}, by converting each of their rules, in a uniform manner, into a symbolic form. 
We characterized the relationship between the two semantics by exhibiting a simulation relation between them. Next, we defined the bounded invariant checking problem for STS over symbolic configurations, as induced for a given \stateflow{} program by the set of the SSOS operational rules, and presented informally a procedure for deriving the initial and next state predicates of the STS. Finally, we showed how to generate, from the STS, a set of quantifier-free FOL assertions in SMT-LIB format suitable for analysis using state-of-the-art SMT solvers. The main benefit of our work is that it lays down the foundations for
the development of tools for the scalable verification of complex industrial \stateflow{} models by means of existing symbolic techniques, which we demonstrated with bounded invariant checking on the running \stateflow{} program example for different values of the depth bound~\emph{k}. We showed that even though in its current state the tool cannot exhaustively verify \stateflow{} models, it still represents a viable choice for fast detection of design errors within a traces of finite length. Even though we initially planned to compare our approach against the state-of-the-practice SLDV tool, we had to withdraw from our idea once we discovered the license constraints imposed by Mathworks. Unfortunately, this happened in the final stage of this work, when it was already too late to steer the validation into a different direction.

Currently, we are working on automating the remaining parts of our technique, and packaging it into a user-friendly and easy to use tool-set. Our goal is to fully automate the transformation of the \stateflow{} programs into a symbolic representation, to be used as input to (bounded) invariant checkers, either applied to \stateflow{} programs in isolation, or as part of already existing frameworks for symbolic analysis of Simulink models (see, e.g., \cite{filipovikj2019boundedinvariancechecking}). Once we have fully developed the tool-set, we will perform a more extensive evaluation of the approach, and compare its performance with that of state-of-the-art tools implementing alternative verification techniques.
Further, we plan to extend our formal characterization of the SSOS in terms of a stronger equivalence between the concrete and symbolic representations of \stateflow{} programs, to formally underpin the symbolic verification of a wider class of properties than invariant properties, such as LTL properties. Finally, we will explore the possibility of extending our BMC approach from refutation-based to a verification one, by adding induction~\cite{demoura03BMCrefutationtoverification}. Along this line of research, we plan to include the option of converting the generated STS into an input format for tools that implement more sophisticated model checking algorithms, such as Lustre~\cite{pilaud1987lustre} models for the Kind2~model~checker~\cite{champion16kind2}. 

\paragraph{Acknowledgements.}

This work has been funded by the Swedish Governmental Agency for Innovation Systems (VINNOVA) under the AVerT project 2018-02727. The authors would also like to thank Gustav Ung for his valuable comments and constructive feedback.


\newpage
\appendix
\section{Complete Set of SSOS Rules}
\label{sec:appendix-A}
\begin{figure}[t!]
\centering
\input{sections/sos-rules/transition-rules}
\caption{SSOS rules for transitions.}
\label{fig:t-rules-complete}
\end{figure}

In this appendix, we present the complete set of SSOS rules. The rules are divided in 5 categories, as follows: rules for transitions (Figure~\ref{fig:t-rules-complete}), transition lists (Figure~\ref{fig:T-rules-part-1}), state definitions (Figure~\ref{fig:sd-rules-complete}), \texttt{And-}compositions  (Figure~\ref{fig:and-rules-complete}), and \texttt{Or-}compositions (Figure~\ref{fig:or-rules-part-1}, and Figure~\ref{fig:or-rules-part-2}).

The set of SSOS rules for \emph{transitions} is composed of 3~rules and is given in Figure~\ref{fig:t-rules-complete}. Intuitively, the rules describe the following. The [t-FIRE] rule describes how a \stateflow{} transition~$t$ fires by appending the symbolic evaluation of the condition $t.c$ to the current path condition and by symbolically executing the condition action $t.ca$ over the current symbolic environment $\Delta$. When a transition fires, a transition event $Fire(t.d, t.ta)$ is generated. The [t-NOT-ENABLED] rule is applicable when the event of the currently processed transition does not match the currently active event. Finally, the last rule [t-NO-FIRE] describes how a \stateflow{} transition does not fire because its guard does not evaluate to true. The processing of such transitions results in a new path condition, which is generated by appending the negation of the transition condition to the current path condition, while the symbolic environment remains the same.

\begin{figure}[ht!]
\centering
    \input{sections/sos-rules/transition-list-rules}
    \caption{SSOS rules for transition list (part 1).}
  \label{fig:T-rules-part-1}
\end{figure}

Next comes the set of SSOS rules for \emph{transitions lists}~$T$, which is given in Figure~\ref{fig:T-rules-part-1}. The list is composed of 7~rules, as follows. [T-$\emptyset$] describes how an empty transition list is processed, rules [T-FIRE], [T-NO-LAST-1], [T-NO-LAST-2], and [T-NO], describe the sequential execution of the constituent transitions of the transition list. If a transition from a transition list fires to a junction ($j$), the list of transitions of that junction is processed in the following manner: the instantaneous processing of transitions until a transition is completed ([T-FIRE-J-F]), a final junction is reached ([T-END]) or the evaluation fails in which case the execution backtracks to resume using the [T-FIRE-J-N] rule.

\begin{figure*}[t!]
\centering
    \input{sections/sos-rules/sd-rules}
    \caption{SSOS rules for state definition.}
  \label{fig:sd-rules-complete}
\end{figure*}

\begin{figure*}[t!]
\centering
\input{sections/sos-rules/and-rules}
\caption{SSOS rules for \texttt{And} composition.}
\label{fig:and-rules-complete}
\end{figure*}

The SSOS rules for \emph{state definitions} (5 in total) are given in Figure \ref{fig:sd-rules-complete}. The first rule, named [SD-NO], is applied when there is no enabled transition for the state definition, be it internal or output. The rule [SD-INT-FIRE] applies when some internal transition within a state definition fires, whereas [SD-FIRE] captures the behavior of the state definition when one of its outgoing transitions fires. [SD-INIT] represents the initialization behavior of the state definitions, and finally [SD-EXIT] describes the sequence of actions when a state definition is exited, when the state definition is placed inside another state.

The set of SSOS rules for \texttt{And}-\emph{compositions} is given in Figure~\ref{fig:and-rules-complete}, and contains the rules for updating ([AND]), initialization ([AND-INIT]) and exiting ([AND-EXIT]), respectively. 

Finally, the set of SSOS rules for \texttt{Or}-compositions is given in Figure~\ref{fig:or-rules-part-1}, and Figure~\ref{fig:or-rules-part-2}, respectively. Rules [OR-EXT-FIRE], and [OR-EXT-FIRE] describe the situations where a transition within an \texttt{Or}-composition fires and the destination is within and outside the composition, respectively. The [OR-NO] rule describes a situation when no transition of an \texttt{Or}-composition fires, whereas the [OR-INT-FIRE] and [OR-FIRE] rules are applicable when an internal or external transition fires, respectively. The [OR-INT-NO-STATE] rule describes how an \texttt{Or}-composition without any internal components fires. The initialization of an \texttt{Or}-composition in situations when the composition enters the initial state, or an incoming transition comes into non-initial internal state, is handled by the [OR-INIT-$\emptyset_p$] and [OR-INIT] rules. Finally, the [OR-EXIT] rule describes the steps involved in exiting from an \texttt{Or}-composition. 

\begin{figure*}[t]
    \centering
    \input{sections/sos-rules/or-rules}
    \caption{SSOS rules for \texttt{Or} composition (part 1).}
    \label{fig:or-rules-part-1}
\end{figure*}

\begin{figure*}[t]
    \centering
    \input{sections/sos-rules/or-rules-2}
    \caption{SSOS rules for \texttt{Or} composition  (part 2).}
    \label{fig:or-rules-part-2}
\end{figure*}
\newpage
\newpage

\end{document}